\documentclass[journal]{IEEEtran}

\ifCLASSINFOpdf
\else
   \usepackage[dvips]{graphicx,color}
\fi
\usepackage{url}

\usepackage{graphicx,balance}
\usepackage{bm}
\usepackage{amssymb}
\usepackage{amsmath}
\usepackage{textcomp}
\usepackage{color}
\usepackage{epstopdf}
\usepackage{subfigure}
\usepackage{booktabs}

\begin{document}

\title{3D  Scene Based Beam Selection for mmWave Communications}

\author{Weihua Xu, Feifei Gao, Shi Jin, and Ahmed Alkhateeb
\thanks{W. Xu and F. Gao are with Institute for Artificial Intelligence Tsinghua University (THUAI),  Beijing, 100084, P. R. China. email: xwh19@mails.tsinghua.edu.cn, feifeigao@ieee.org)}
\thanks{S. Jin is with the National Mobile Communications Research Laboratory,
Southeast University, Nanjing 210096, China (e-mail: jinshi@seu.edu.cn)}
\thanks{A. Alkhateeb is with the Department of Electrical and Computer Engineering, University of Texas at Austin, Austin, TX 78712-1687 USA (e-mail:
aalkhateeb@utexas.edu)}}

%\markboth{Journal of \LaTeX\ Class Files, Vol. 14, No. 8, August 2015}
\maketitle

\begin{abstract}
In this paper, we present a novel framework of 3D scene based beam selection for mmWave communications that relies only on the environmental data and deep learning techniques. Different from other \emph{out-of-band side-information} aided communication strategies, the proposed one fully utilizes the environmental information, e.g., the shape, the position, and even the materials of the surrounding buildings/cars/trees that are obtained from 3D scene reconstruction. Specifically, we build the neural networks with the input as point cloud of the 3D scene and the output as the beam indices. Compared with the LIDAR aided technique, the reconstructed 3D scene here is achieved from multiple images taken offline from cameras and  thus significantly lowers down the cost and makes itself applicable for small mobile terminals. Simulation results show that the proposed 3D scene based beam selection can outperform the LIDAR method in terms of accuracy.

\end{abstract}


\IEEEpeerreviewmaketitle

\section{Introduction}

Millimeter wave (mmWave) communication is an essential part of 5G due to its large bandwidth of radio spectrum. Generally,  massive number of  antennas are adopted at the base station (BS) and the narrow beamforming is formulated to overcome the high path attenuation of mmWave band and to achieve ultra high transmission rate. However, narrow beam-width makes the beam alignment a difficult task, especially in high mobility scenarios like vehicle-to-everything (V2X)  communications. In this case, the traditional beamforming strategies, such as beam sweeping or beam computing, that rely on channel estimation would be repeated frequently and thus bring significant time and energy overhead. Hence, research on highly efficient beamforming schemes to avoid latency caused by beam misalignment or beam switching has attracted increasing attention in recent years \cite{Ahmed}-\cite{Giordani}.

Using \emph{out-of-band side-information} currently becomes one major approach to reduce channel estimation or beam selection overhead. In \cite{wang1}, the out-of-band spatial information obtained from sub-6 GHz is used for mmwave beam selection. In \cite{d2dspectrum}-\cite{wanga}, the fingerprinting-based database that is composed of the side-information and the corresponding beam information are used for beam selection, where the side-information can be vehicle positions obtained by the Global Positioning System (GPS). In \cite{zhou}, the positions of multiple vehicles are adopted through the machine learning tools under the vehicle-to-infrastructure (V2I) scenario. As vehicles can equip  auxiliary sensors \cite{xing1}, such as radar, LIDAR, camera, etc., a more recent work utilizes the point cloud scanned by LIDAR to improve the efficiency of beam selection \cite{xing2}-\cite{tayyaba}, which is shown to have higher accuracy than the position-aided method \cite{tayyaba}. However, using the expensive high precision LIDAR may not be always available and affordable, especially for portable terminals like mobile phones. Hence, some other affordable ways to get point clouds should be explored.

In this paper, we present a new framework of how to utilize the surrounding 3D scene of the cellular coverage to help the wireless communications and to reduce the training overhead. Specifically, we propose a novel 3D scene based beamforming selection method by utilizing panoramic scene feature. The 3D scene reconstruction technique works by taking images from ordinary cameras for cellular coverage area from many perspectives. Then the panoramic point cloud can be built and is stored at the base station or at the moble station (MS). Such an approach does not have the huge cost as compared to the LIDAR-based method and is much cheaper for mobile users or BSs. With the information of MS's position, we design a panoramic scene feature from the saved point cloud to represent the relative spatial relationship between BS and MS. Then, such panoramic scene feature can be utilized to predict the optimal transmit and receive beams simultaneously through training a 3D deep neural network (DNN). Furthermore, the proposed 3D scene based approach does not require real-time LIDAR scanning for acquiring local scene
information compared with the LIDAR methods \cite{xing2}-\cite{tayyaba}. While compared with the LIDAR distributed method \cite{xing2}, this 3D scene based approach can adapt to new environments in a better way and achieve a higher beam prediction accuracy, as will be seen in the simulation part.
\begin{figure*}[t]
\centering
\includegraphics[width=1\textwidth]{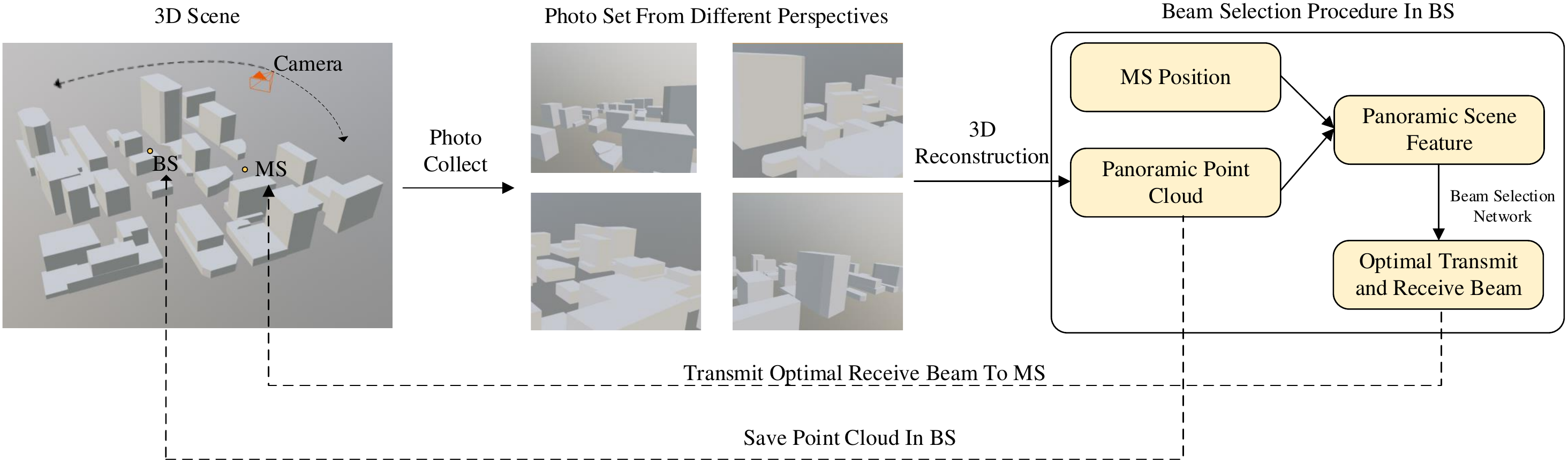}
\caption{The proposed framework for 3D scene based beam selection.}
\end{figure*}

\section{System Model}
Let us consider a downlink mmWave Multi-input Multi-output (MIMO) system with a single MS, while this 3D scene based framework can be straightforwardly adapted to multi-user scenario. BS is equipped with a uniform planar array (UPA) of $N_\mathrm{B}=N_\mathrm{B}^{\mathrm{a}}\times N_\mathrm{B}^{\mathrm{b}}$ antennas, and MS is equipped with a UPA of $N_\mathrm{M}=N_\mathrm{M}^{\mathrm{a}}\times N_\mathrm{M}^{\mathrm{b}}$ antennas. The antenna spacing of transmit and receive UPA is half carrier wavelength. We assume BS and MS both have a single radio frequency (RF) chain at mmWave band to save the cost. The downlink signal received at the user can be expressed as
\begin{equation}
y=\bm{\mathrm{f}}^{\mathrm{H}}_{\mathrm{M}}\bm{H}\bm{\mathrm{f}}_{\mathrm{B}}s+\bm{\mathrm{f}}^{\mathrm{H}}_{\mathrm{M}}\bm{n},
\end{equation}
where $\bm{H}\in \mathbb{C}^{N_\mathrm{M}\times N_\mathrm{B}}$ is the channel matrix, $\bm{\mathrm{f}}_{\mathrm{M}}\in \mathbb{C}^{N_\mathrm{M}\times 1}$ is the receive beamforming vector, $\bm{\mathrm{f}}_{\mathrm{B}}\in \mathbb{C}^{N_\mathrm{B}\times 1}$ is the transmit beamforming vector, and $\bm{n} \in \mathcal{CN}(0,\sigma^2\bm{\mathrm{I}})$ is the Gaussian noise signal. For mmWave transmission, the geometric channel model is usually adopted and the channel matrix $\bm{H}$ is expressed as
\begin{equation}
  \bm{H}=\sum_{l=1}^{N_\mathrm{p}}\alpha_l\bm{a}_\mathrm{r}(\theta^r_l,\phi^r_l)\bm{a}_\mathrm{t}^{\mathrm{H}}(\theta^t_l,\phi^t_l),
\end{equation}
where $\alpha_l$ is the complex gain of the $l$th path, $\theta^\mathrm{r}_l$ and $\phi^\mathrm{r}_l$ are the elevation and azimuth of the $l$th path's angle of arrival (AOA), $\theta^\mathrm{t}_l$ and $\phi^\mathrm{t}_l$ are the elevation and azimuth of the $l$th path's angle of departure (AOD), $\bm{a}_\mathrm{r}(\theta,\phi) \in \mathbb{C}^{N_\mathrm{M}\times 1}$ and $\bm{a}_\mathrm{t}(\theta,\phi) \in \mathbb{C}^{N_\mathrm{B}\times 1}$ are the complex steering vectors of receive and transmit array. In addition, the steering vector of a UPA with $N^\mathrm{a} \times N^\mathrm{b}$ antennas is given by
\iffalse
\begin{equation}
\bm{a}(N^\mathrm{a},N^\mathrm{b},\theta,\phi)=\frac{1}{\sqrt{N^{\mathrm{a}}N^{\mathrm{b}}}}\bm{a}_{\mathrm{az}}(\theta,\phi)\bigotimes\bm{a}_{\mathrm{ele}}(\theta),
\end{equation}
where
\begin{align}
&\bm{a}_{\mathrm{ele}}(\theta)=[1,e^{j\pi \cos(\theta)},\cdots,e^{j(N^{\mathrm{a}}-1) \pi \cos(\theta)}]^{\mathrm{T}},\\
&\bm{a}_{\mathrm{az}}(\theta,\phi)=
\begin{aligned}
&[1,e^{j\pi \sin(\theta)\sin(\phi)},\\
&\cdots,e^{j(N^{\mathrm{b}}-1) \pi \sin(\theta)\sin(\phi)}]^{\mathrm{T}},
\end{aligned}
\end{align}
\fi
\begin{equation}
\begin{aligned}
\bm{a}(N^\mathrm{a},N^\mathrm{b},&\theta,\phi)=\frac{1}{\sqrt{N^{\mathrm{a}}N^{\mathrm{b}}}}[1,\cdots,e^{j\pi[w \cos(\theta)+h\sin(\theta)\sin(\phi)]}\\
&\cdots,e^{j\pi[(N^{\mathrm{a}}-1) \cos(\theta)+(N^{\mathrm{b}}-1)\sin(\theta)\sin(\phi)]}]^{\mathrm{T}},
\end{aligned}
\end{equation}
and $0\leq w\leq N^{\mathrm{a}}-1$,$0\leq h\leq N^{\mathrm{b}}-1$. Then, we have $\bm{a}_\mathrm{r}(\theta,\phi)=\bm{a}(N^\mathrm{a}_\mathrm{M},N^\mathrm{b}_{\mathrm{M}},\theta,\phi)$ and $\bm{a}_\mathrm{t}(\theta,\phi)=\bm{a}(N^\mathrm{a}_\mathrm{B},N^\mathrm{b}_\mathrm{B},\theta,\phi)$.

The codebook-based beamforming strategy is generally adopted in practice, where the pair of beamforming vectors ($\bm{\mathrm{f}}_{\mathrm{M}}$,$\bm{\mathrm{f}}_{\mathrm{B}}$) should be selected from the preset transmit beam codebook $\bm{\mathfrak{F}}_{\mathrm{B}}=\{\bm{\mathrm{f}}_{\mathrm{B},1},\bm{\mathrm{f}}_{\mathrm{B},2},\cdots,\bm{\mathrm{f}}_{\mathrm{B},n}\}$ and the receive beam codebook $\bm{\mathfrak{F}}_{\mathrm{M}}=\{\bm{\mathrm{f}}_{\mathrm{M},1},\bm{\mathrm{f}}_{\mathrm{M},2},\cdots,\bm{\mathrm{f}}_{\mathrm{M},m}\}$.
The target is to maximize the   receive signal-to-noise ratio (SNR)  and thus selects the optimal beamforming vector pair from \begin{equation}
(\bm{\mathrm{f}}_{\mathrm{M}}^{\mathrm{opt}},\bm{\mathrm{f}}_{\mathrm{B}}^{\mathrm{opt}})=\mathop{\arg\max}_{\bm{\mathrm{f}}_{\mathrm{M}} \in \bm{\mathfrak{F}}_{\mathrm{M}}, \bm{\mathrm{f}}_{\mathrm{B}} \in \bm{\mathfrak{F}}_{\mathrm{B}}} |\bm{\mathrm{f}}^{\mathrm{H}}_{\mathrm{M}}\bm{H}\bm{\mathrm{f}}_{\mathrm{B}}|^2/\|\bm{\mathrm{f}}_{\mathrm{M}}\|^2. \label{gao:1}
\end{equation}

\section{The 3D Scene based Beamforming}

To maximize (\ref{gao:1}), the conventional communications approach needs first obtain $\mathbf{H}$ and then BS needs to send a large amount of training signals, which costs much spectral resources, especially when the BS has multiple antennas. In this work, we propose to reconstruct the 3D scene with the images obtained from the camera at BS and MS or from any other means, and then relate it with parameters in wireless communications, e.g., the optimal beam index. Interestingly, 3D scene reconstruction is a core problem in the field of computer vision. We here explore the roles of 3D reconstruction techniques for wireless communication, since it can represent the full spatial feature between BS and MS. It is expected that 3D scene based approach to outperform the conventional candidates, such as LIDAR methods, that only leverage partial spatial information from sensors.

Fig.~1 illustrates the proposed framework with an example of BS storing the point cloud. We first take enough photos\footnote{The photos can be obtained from MSs, unmanned aerial vehicles (UAVs), or any other approaches.} offline for the cellular coverage. Next we apply the 3D reconstruction methods to rebuild the 3D scene point cloud of the whole cell. When an MS needs to establish the mmWave communication link with BS, it could report its position, and then BS can construct a panoramic scene feature by combining MS's position coordinates and the panoramic point cloud. Such a panoramic feature contains rich spatial information of the channel between BS and MS, such as the number, the positions and sizes of buildings, and will be used as the input feature of the designed 3D DNN. The output of this 3D DNN is deemed as the optimal beam label. Hence, once the DNN is pre-trained well by large number of samples stored in BS, the BS can utilize the DNN and the panoramic scene feature to predict the optimal beam without beam training. In cases that the MS's position is private, the panoramic point cloud of the cell can be downloaded at MS in advance, and MS could pre-save and use the trained DNN by itself and then feedback BS the indices of the selected beams. The above framework is illustrated in details as follows:
\subsection{Panoramic Point Cloud Generation}
With no practical data in hand, we resort to the computer generated data to verify the proposed framework. Specifically, we use Blender \cite{fdma}, a the 3D modeling software, to create several building models and combine them randomly into different environments. An example is shown in Fig. 2, where we choose simple brick texture to overspread the surface of all the buildings to simulate real buildings.

To get an better reconstruction, we choose to reconstruct the point cloud of each building one by one for each scene. Then, the point cloud of the whole scene can be generated by combining all buildings' point cloud based on their locations, which are easy to acquire. This  approach can reduce the difficulty of reconstructing the whole scene and improve the reconstruction accuracy. For each building, we require the camera to revolve around the building at different heights and distances, and keep the camera lens facing the building. After taking photos at designed perspectives, heights and distances, we can obtain an image set of the corresponding environment. The reconstruction is achieved by inputting  the obtained image set to COLMAP \cite{tan}, which is an open-source 3D reconstruction tool based on Structure-from-Motion \cite{relayd2d} and Multi-View Stereo \cite{saleem} algorithms. Then, we can use the reconstructed point clouds of all buildings to reconstruct the whole 3D scene.

\subsection{Panoramic Scene Feature Extraction}
\begin{figure}[t]
\centering
\includegraphics[width=0.43\textwidth]{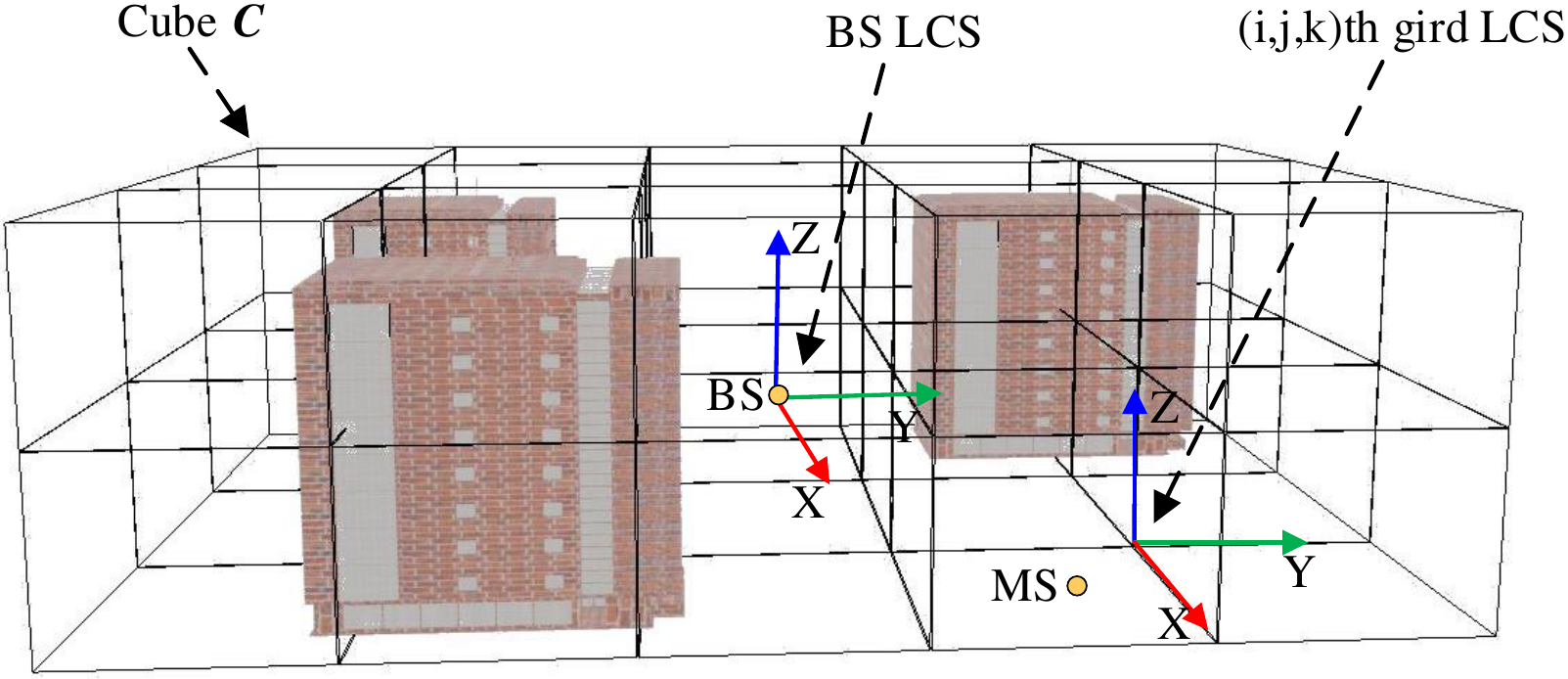}
\caption{ One 3D environment and the BS/Grid local coordinate system}
\end{figure}
We can express the reconstructed point cloud of a generated 3D scene as a set $\bm{P}=\{\bm{p}_1,\bm{p}_2,\cdots,\bm{p}_{K}\}$, where $K$ denotes  the number of all points in the point cloud and each element $\bm{p}_k \in \mathbb{R}^{3\times 1}$  represents the three-dimensional coordinates of one point. The BS is assumed to already know its absolute coordinate $\bm{p}_\mathrm{B}\in \mathbb{R}^{3\times 1}$ (also known to MS). Hence, when the MS gets its own absolute coordinates $\bm{p}_\mathrm{M}\in \mathbb{R}^{3\times 1}$ and feedbacks to BS, BS or MS can construct a panoramic scene feature containing both their relative spatial relationship and BS's surrounding environment by using $\bm{P}$, and $\bm{p}_\mathrm{M}-\bm{p}_\mathrm{B}$.

As BS have limited coverage, we can find a suitable specific cube area $\bm{C}$ that remains fixed under BS's local coordinate system (LCS) and contain all or most of the panoramic point cloud of the BS¡¯s surrounding environments, as Fig.~2 shows. We evenly divide the length, the width and the height of $\bm{C}$ into $a$, $b$ and $c$ parts respectively, and divide $\bm{C}$ into $a*b*c$ small cubic blocks. The core idea of designing the panoramic feature is to use the small cubic blocks to segment the point cloud $\bm{P}$, and set a LCS for each cubic grid. For each grid, the origin of the corresponding LCS is set to be one vertex of this grid. The LCSs of all grids are parallel, and each grid is in the first quadrant of the corresponding LCS. Hence, the local coordinates of each point in $\bm{P}$ will be positive under the LCS of the grid containing this point. Then, we can partition the $K$ points into $a*b*c$ spatial blocks to construct the panoramic scene feature, named as $\bm{g} \in\mathbb{R}^{a\times b\times c\times 3} $, where the $\bm{g}_{[i,j,k,:]}$ is the mean value of local coordinates of those points only appearing in the $(i,j,k)$th spatial block. We can set one row of $\bm{g}$ be $\bm{0}$ if there are no points in the corresponding spatial grid.

Since we need to further add the information to represent relative spatial relationship between BS and MS to the panoramic scene feature $\bm{g}$, we will compute MS's local coordinate $\tilde{\bm{p}}_{\mathrm{M}}$ under the LCS of the $(i_{\mathrm{M}},j_{\mathrm{M}},k_{\mathrm{M}})$th grid, which contains MS, based on the value of $\bm{p}_\mathrm{M}-\bm{p}_\mathrm{B}$. Then, we can set $\bm{g}_{[i_{\mathrm{M}},j_{\mathrm{M}},k_{\mathrm{M}},:]}=-\tilde{\bm{p}}_{\mathrm{M}}$ to distinguish MS from the point cloud $\bm{P}$, since we express the MS's local coordinates by negative numbers in $\bm{g}$. Afterwards, the panoramic scene feature $\bm{g}$ can represent the spatial information of the channel between BS and MS and will be used as the input feature of the designed DNN. Compared with the original point cloud $\bm{P}$, the dimension of $\bm{g}$ is significantly reduced while such feature $\bm{g}$ still works efficiently, as will be seen in the later simulations.

\subsection{Deep Learning Model for Beam Selection}
\begin{figure}[t]
\centering
\includegraphics[width=0.5\textwidth]{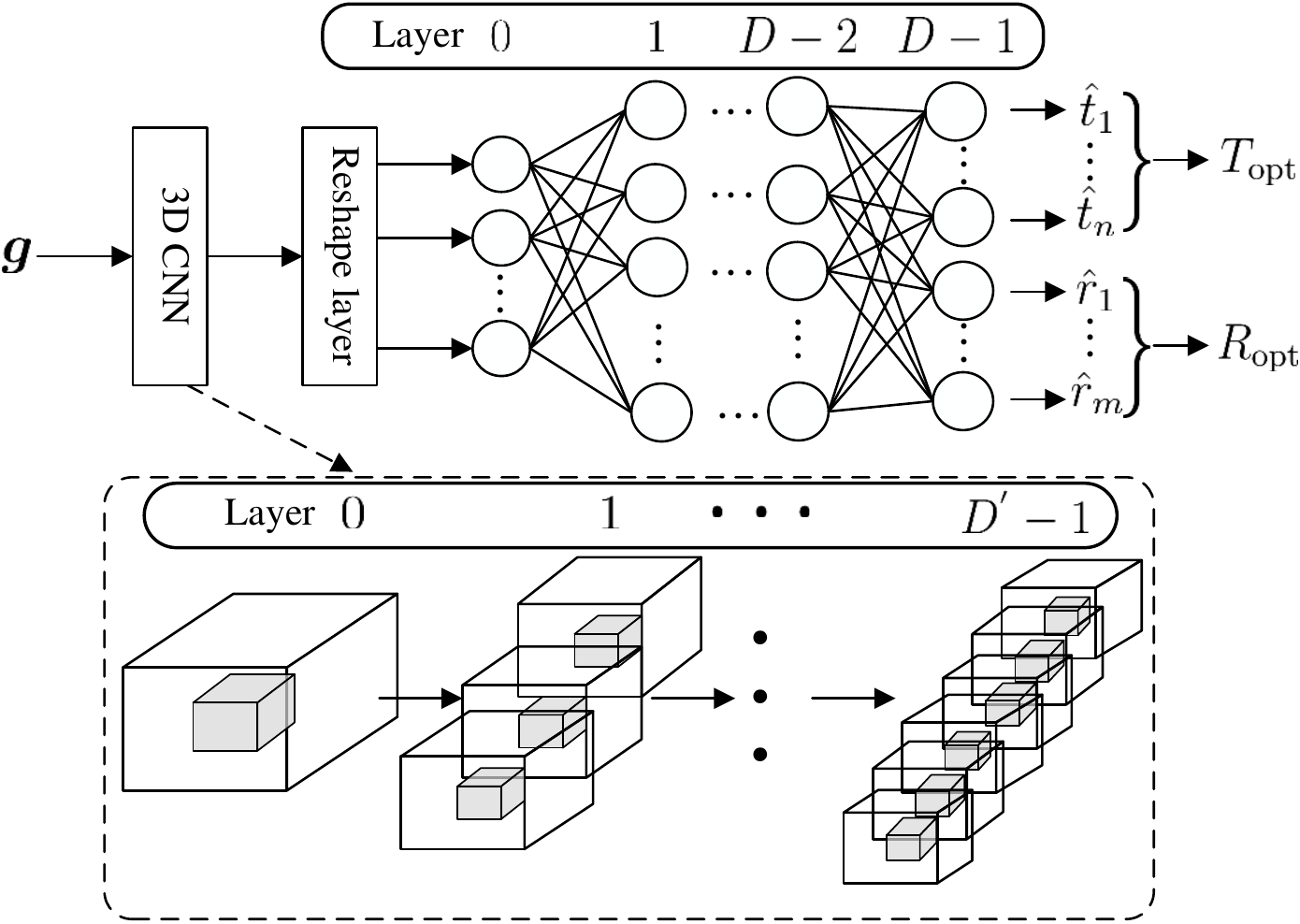}
\caption{The proposed deep learning network structure for beam selection}
\end{figure}
Fig.~3 shows the proposed deep learning network with the 3D convolutional neural network (CNN), where the inputs are the panoramic scene features and the outputs are the optimal transmit beam index $T_\mathrm{opt}$ and the optimal receive beam index $R_\mathrm{opt}$. Our target is to design a general DNN that can adapt to different environments, and the detailed training phase is explained as follows:

We generate numerous environments in a
random way and select the positions of MS for each generated environment. For each environment and each MS postion, the corresponding channel is produced by the ray tracing software Wireless Insite \cite{tdma}. Thus, the panoramic scene feature and the corresponding optimal transmit/receive beam indices can be acquired at each position of MS to build a training sample. The total training samples built in each generated environment  formulate the training set. Then we train the DNN that predicts the optimal transmit/receive beamforming vectors by utilizing the RMSProp optimizer and the cross entropy loss function.

\section{Simulation Results}
\subsection{Simulation Setup}
\begin{figure}[t]
\centering
\includegraphics[width=0.5\textwidth]{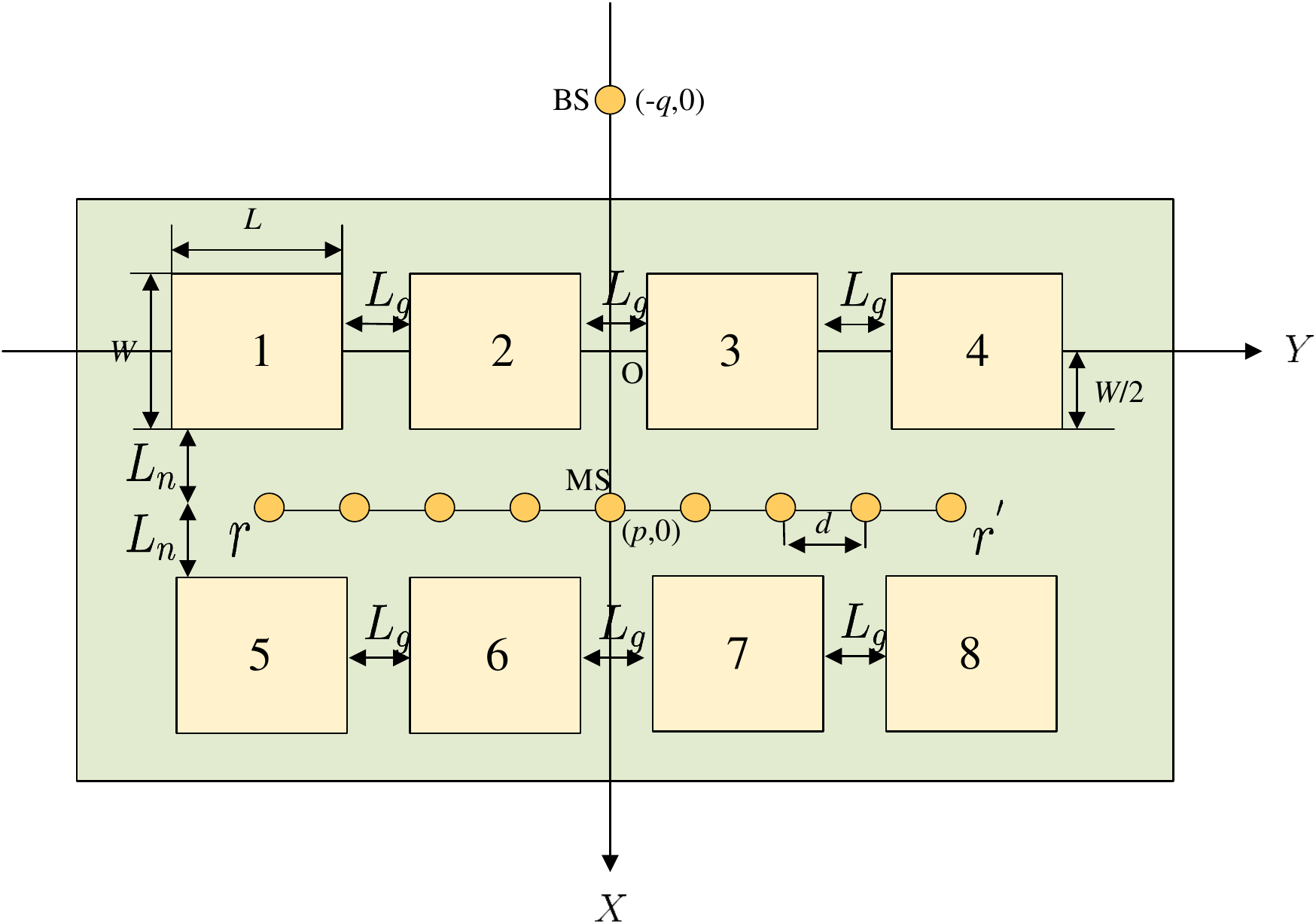}
\caption{The top view of BS coverage area for the simulated environments}
\end{figure}

We design three types of cube-like building models and summarise their sizes in TABLE $\textrm{I}$. The random way of generating environments for simulation is explained in Fig.~4, whcih is the top view of BS coverage area. We focus on the coverage area, i.e., the largest rectangular in Fig.~4 and select eight square areas (SA) indexed 1 to 8 with the same size. When creating a new environment, five SAs will be selected by random extraction with equal probability. In each SA, there will be a building whose type and position are also generated randomly with uniform distribution, and only building positions that can ensure the occupied area of the building does not exceed
the SA will probably be selected. The above approach could ensure that the randomly generated buildings do not overlap.

\begin{table}[t]
\centering
\caption{Building Sizes For Simulation}
\begin{tabular}{cccc}
\toprule
Type& Length/m& Width/m& Height/m\\
\midrule
Type A& 25.6& 18.5& 25\\
Type B& 20& 14& 25\\
Type C& 12& 9& 15\\
\bottomrule
\end{tabular}
\end{table}

\begin{table}[t]
\centering
\caption{Critical Parameters Of Wireless Insite For Ray Tracing}
\begin{tabular}{cccc}
\toprule
Parameter& Value\\
\midrule
Carrier Frequency& 60 $\mathrm{GHz}$\\
Propagation Model& X3D\\
Building Material& Concrete\\
Diffuse Scattering Reflections& 2\\
Diffuse Scattering Diffractions& 1\\
Maximum Paths Per Receiver Point& 25\\
\bottomrule
\end{tabular}
\end{table}

The positions of MS in each environment are  evenly spaced on the fixed line segment $rr^{'}$, as illustrated in Fig. 4. The parameters to generate the environments are $W=36.5\mathrm{m}$, $L=35.6\mathrm{m}$, $L_g=5 \mathrm{m}$, $rr^{'}=80 \mathrm{m}$, $d=0.5\mathrm{m}$, $p=60\mathrm{m}$, and $q=45\mathrm{m}$, respectively. The height of BS and MS is set to $10 \mathrm{m}$ and $2 \mathrm{m}$ respectively. Thus we can obtain 161 training samples for each environment. We generate $300$ environments with the restriction that area 2 or area 3 must have buildings, i.e., there is no simple line-of-sight scenario. All environments' point cloud are reconstructed as described in Section III.B. The cube area $\bm{C}$ of all environments are kept the same with length $200 \mathrm{m}$, width $160 \mathrm{m}$, and height $30 \mathrm{m}$. We set $a,b,c$ to $40,32,6$ respectively. Thus, we get 38640 samples obtained in 240 environments as the training set, and get 9660 samples obtained in other 60 environments as the testing set.

The main parameters of Wireless Insite to generate channels are described in TABLE. $\textrm{II}$. The antenna numbers of arrays in BS and MS are $N_\mathrm{B}^{\mathrm{a}}=N_\mathrm{M}^{\mathrm{a}}=8$ and $N_\mathrm{B}^{\mathrm{b}}=N_\mathrm{M}^{\mathrm{b}}=72$, respectively. We adopt the beam codebook with equal azimuth angle interval. We let BS and MS both have the same codebook with $N=30$ and $N=50$ beams i.e., $\bm{\mathrm{f}}_{\mathrm{B},i}=\bm{\mathrm{f}}_{\mathrm{M},i}= \bm{a}(N_\mathrm{B}^{\mathrm{a}},N_\mathrm{B}^{\mathrm{b}},\hat{\alpha},\frac{2i-2-N}{2N}\pi)$, $i=1,2,\cdots,N$, where $N$ is number of total beams in codebook and $\hat{\alpha}$ is fixed to $95^{\circ}$ according to the horizontal line constraint for MS position and the BS's/MS's heights.

We set $D=6$ and $D^{'}=2$ for our deep learning network. For the 3D CNN, the kernel size, number of filters and strides of layer 0 are set to be $(5,5,2)$, $6$ and $(2,2,1)$ respectively. The kernel size, number of filters and strides of layer 1 is set to be $(3,3,2)$, $12$ and $(2,2,1)$ respectively. For the fully-connected network after the reshape layer, the node number of each layer except the output layer is set to be $2304$, $1600$, $1000$, $500$, $100$ respectively in index order.

\subsection{Results and Discussions}
As shown in Fig.~5, we analyze the top-$5$ selection accuracy of the proposed method under different sizes of the training set.  With more training samples as input, the selection accuracy of proposed method increases. When more than 8000 samples are used as input, the increasing of the selection accuracy starts to slow down. In this case, one may try to find other ways to further increase the accuracy, such as modeling based approach or a few pilot signals.

\begin{figure}[t]
\centering
\includegraphics[width=0.5\textwidth]{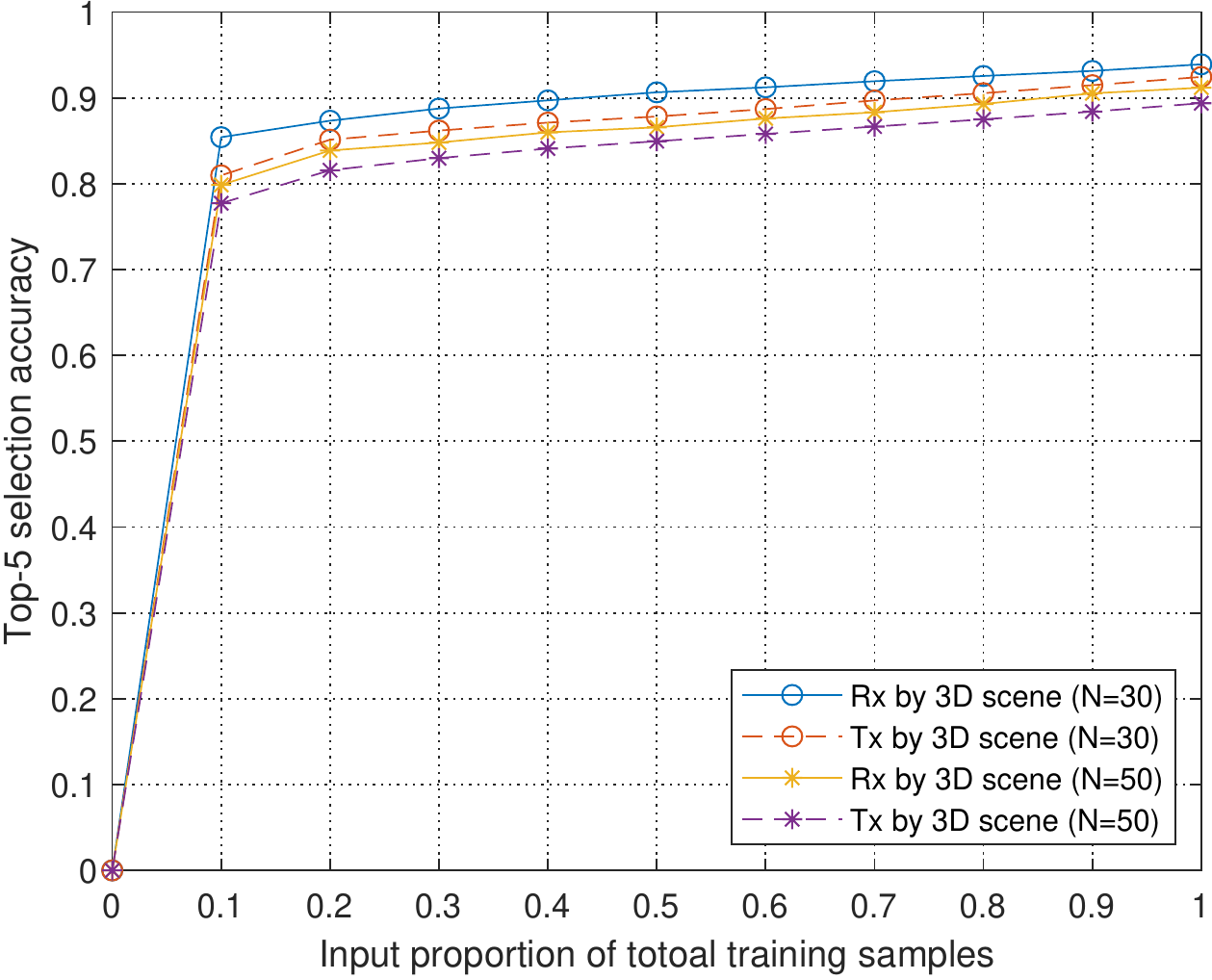}
\caption{Top-$5$ beam selection accuracy of the  proposed method with different input training set sizes}
\end{figure}
\begin{figure}[t]
\centering
\includegraphics[width=0.5\textwidth]{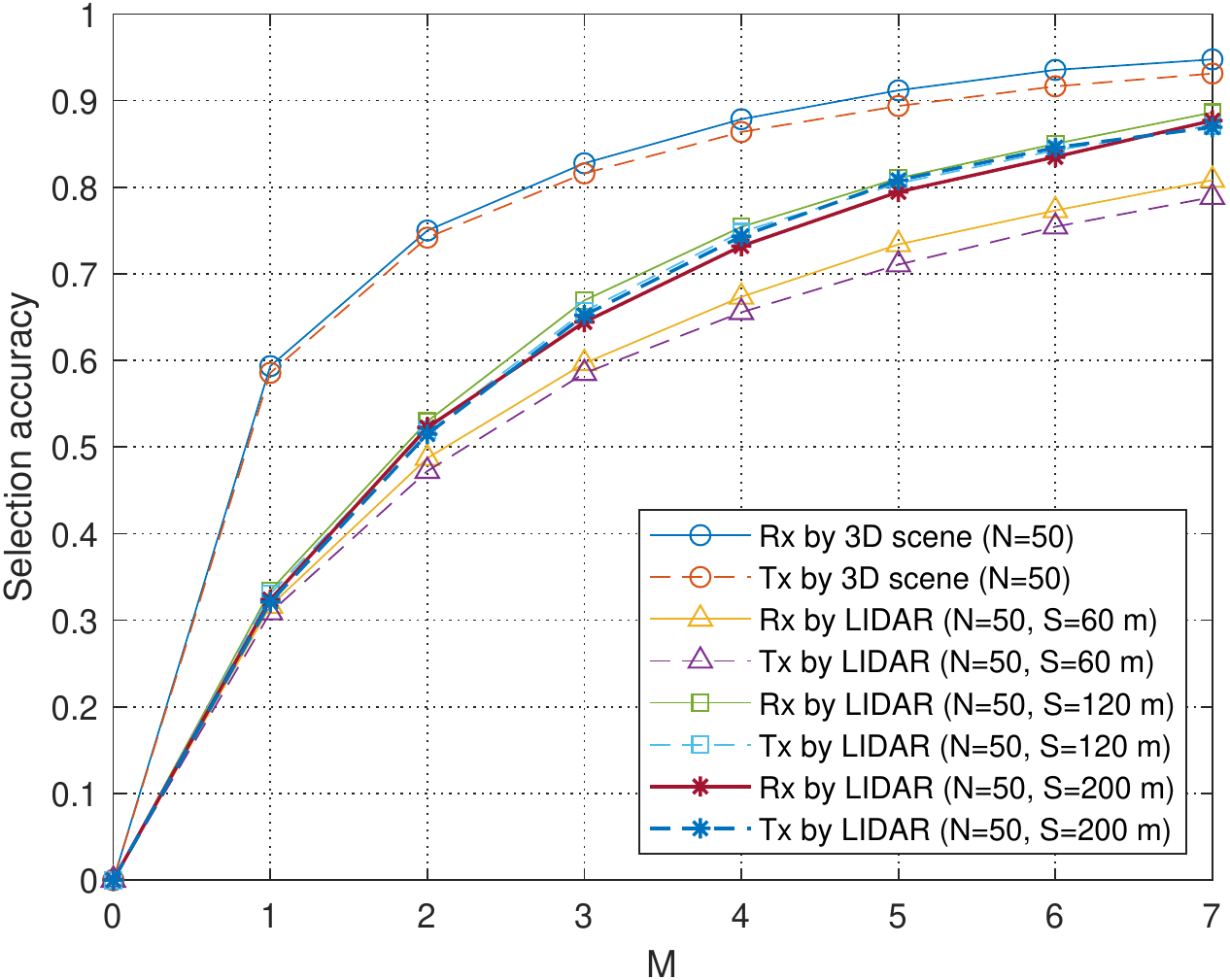}
\caption{Top-$M$ beam selection accuracy of proposed method and the LIDAR distributed method}
\end{figure}
We next compare the proposed method with the LIDAR distributed method [9], which performs the scanning point cloud and predicts the optimal beam pair at MS. For a fair comparison, the two methods utilize the same design of neural network that is indicated in Section IV.A, and the input feature design of the compared LIDAR method is consistent with the one in [9]. As shown in Fig.~6, the performance of this LIDAR method are simulated for different LIDAR observational distances $S$ that are set as $60 \mathrm{m}$, $120 \mathrm{m}$ and $200 \mathrm{m}$ respectively. It is seen that the performance of the LIDAR method with observational distance $120 m$ or $200 m$ is better than the LIDAR method with  observational distance $60 m$, and the performance is almost the same for  observational distance $120 m$ and $200 m$. The results indicate $120 m$ and $200 m$ are all long enough to scan the whole BS coverage area for acquiring sufficient local spatial information. Moreover, the performance of the LIDAR method can not be further improved by increasing observational distance after $200 m$. It is also seen that the proposed method can outperform the LIDAR distributed method with observational distance as $200 m$ by $20\%$ on average and up to nearly $30\%$ in top-$M$ selection accuracy. Hence, we can get the conclusion that the panoramic point clouds constructed by the proposed method are more appropriate for accurate beam selection when compared with the local point cloud scanned by single MS's LIDAR. The results can be explained by the richer spatial information of the proposed panoramic point cloud than the part spatial information of single LIDAR scaned point cloud. Another reason is that the proposed method have a better generalization performance for new scenes, whereas the LIDAR distributed method works well for single environment \cite{tayyaba}. It needs to emphasized that the energy and time overhead of the proposed 3D scene reconstruction method is much lower than LIDAR distributed method, because the proposed method only needs to performing 3D reconstruction once for each scene and does not require real-time LIDAR scanning.

\section{Conclusions}
We have proposed a novel beam selection method through image-based 3D reconstruction in order to avoid the severe overhead of using expensive auxiliary devices, such as Radar and LIDAR. With proper training, the proposed approach can predict the optimal beam for any point in the current cell and it can even work for a new environment with the same type of building distributions. Simulations show that the proposed approach has higher beam selection accuracy and lower overhead compared to the LIDAR distributed method, which indicates the advantage of exploring image processing techniques. It is also one of the first attempts to relate the 3D scene and the wireless communications. More applications benefited from 3D reconstruction or other image techniques within wireless communications are of interest and deserve further exploration.

\balance


\begin{thebibliography}{99}

\bibitem{Ahmed}
V. D. P. Souto, R. D. Souza, B. F. Uch\^o-Filho, and Y. Li, ``A novel efficient initial access method for 5G millimeter wave communications using genetic algorithm," {\it IEEE Trans. Veh. Technol.}, vol. 68, no. 10, pp. 9908--9919, Oct. 2019.

\bibitem{Giordani}
M. Giordani, M. Mezzavilla, C. N. Barati, S. Rangan, and M. Zorzi, ``Comparative analysis of initial access techniques in 5G mmWave cellular networks," in {\it Proc. Annu. Conf. Inf. Sci. Syst. (CISS)}, Mar. 2016, pp. 268--273.

\bibitem{wang1}
A. Ali, N. Gonz¨¢lez-Prelcic, and R. W. Heath, ``Millimeter wave beam-selection using out-of-band spatial information," {\it IEEE Trans. Wireless Commun.}, vol. 17, no. 2, pp. 1038--1052, Feb. 2018.


\bibitem{d2dspectrum}
J. C. Aviles, and A. Kouki, ``Position-aided mm-wave beam training under NLOS conditions," {\it IEEE Access}, vol. 4, pp. 8703--8714, 2016.

\bibitem{d2doff}
Z. Wei, Y. Zhao, X. Liu, and Z. Feng, ``DoA-LF: A location fingerprint positioning algorithm with millimeter-wave," {\it IEEE Access}, vol. 5, pp. 22678--22688, 2017.

\bibitem{wanga}
K. Satyanarayana, M. El-Hajjar, A. A. M. Mourad, and L. Hanzo, ``Deep learning aided fingerprint-based beam alignment for mmWave vehicular communication," {\it IEEE Trans. Veh. Technol.}, vol. 68, no. 11, pp. 10858--10871, Nov. 2019.

\bibitem{zhou}
Y. Wang, A. Klautau, M. Ribero, A. C. K. Soong, and R. W. Heath, ``MmWave vehicular beam selection with situational awareness using machine learning," {\it IEEE Access}, vol. 7, pp. 87479--87493, 2019.

\bibitem{xing1}
L. Liang, H. Peng, G. Y. Li, and X. Shen, ``Vehicular communications: a physical layer perspective," {\it IEEE Trans. Veh. Technol.}, vol. 66, no. 12, pp. 10647--10659, Dec. 2017.

\bibitem{xing2}
A. Klautau, N. Gonz¨¢lez-Prelcic, and R. W. Heath, ``LIDAR data for deep learning-based mmWave beam-selection," {\it IEEE Wireless Commun. Lett.}, vol. 8, no. 3, pp. 909--912, June 2019.

\bibitem{tayyaba}
M. Dias, A. Klautau, N. Gonz¨¢lez-Prelcic, and R. W. Heath, ``Position and LIDAR-aided mmWave beam selection using deep learning," in {\it Proc. IEEE Int. Workshop on Signal Processing Adv. in Wireless Commun. (SPAWC)}, Cannes, France, 2019, pp. 1--5.

\bibitem{fdma}
https://www.blender.org

\bibitem{tan}
https://colmap.github.io

\bibitem{relayd2d}
J. L. Sch\"onberger and J. Frahm, ``Structure-from-motion revisited," in {\it Proc. IEEE Conf. on Computer Vision and Pattern Recognition (CVPR)}, Las Vegas, NV, 2016, pp. 4104--4113.


\bibitem{saleem}
J. L. Sch\"onberger, E. Zheng, J. Frahm, and M. Pollefeys, ``Pixelwise view selection for unstructured multi-view stereo,"  in {\it Proc.  Euro. Conf. on Computer Vision (ECCV)}, Amsterdam, OC, 2016, pp. 501--518.


\bibitem{tdma}
https://www.remcom.com/wireless-insite-em-propagation-software




\end{thebibliography}
\end{document}